\documentclass[12pt,preprint]{aastex}

\usepackage{amsmath}
\usepackage{graphicx}
\usepackage{natbib}

\begin{document}

\title{Consolidating and Crushing Exoplanets: Did it happen here?}

\author{Kathryn Volk\altaffilmark{1,2} and Brett Gladman\altaffilmark{1}}
\altaffiltext{1}{Department of Physics and Astronomy, University of British Columbia, 6224 Agricultural Road, Vancouver, BC V6T 1Z1, Canada}
\altaffiltext{2}{CITA National Fellow}

\author{Accepted for publication in ApJ Letters May 26, 2015}

\begin{abstract}

The Kepler mission results indicate that systems of tightly-packed inner planets (STIPs) are present around of order 5\% of FGK field stars (whose median age is $\sim5$~Gyr). We propose that STIPs initially surrounded nearly all such stars and those observed are the final survivors of a process in which long-term metastability eventually ceases and the systems proceed to collisional consolidation or destruction, losing roughly equal fractions of systems every decade in time. In this context, we also propose that our Solar System initially contained additional large planets interior to the current orbit of Venus, which survived in a metastable dynamical configuration for 1-10\% of the Solar System's age.  Long-term gravitational perturbations caused the system to orbit cross, leading to a cataclysmic event which left Mercury as the sole surviving relic.  

\end{abstract}


\section{Absent Planets}\label{s:intro}

Why {\it aren't} there planets interior to Mercury?  This question is particularly evident in light of the discovery of many multi-planet systems at distances $<$0.5~AU containing several Earth masses of material. Our answer is: there were, and Mercury is all that remains, which fits our Solar System into a framework where dynamical instability mercilessly consolidates or degrades close-in planets.

The Kepler mission discovered many systems of tighty-packed inner planets (STIPs)~\citep{Fabrycky2014,Rowe2014,Lissauer2014}.  \citet{Lissauer2011} estimate that $\sim5\%$ of Kepler stars host STIPs and \citet{Fressin2013} conclude that half of the Kepler stars have at least one $0.8-2~R_{\earth}$ planet with orbital periods shorter than Mercury's.  The absence of such close-in planets in our Solar System is perhaps surprising; the surface mass density $\sigma$ profile for the minimum mass solar nebula (with radial dependence $\sigma~\propto~a^{-1}$~or~$a^{-1.5}$) yields several Earth-masses of material inside 0.7~AU if the disk extends down to the $\sim0.05$~AU distance where STIPs are found and where the inner edge of gaseous protoplanetary disks are thought to be \citep{Meyer1997}.  

In contrast, Solar System terrestrial planet formation models require an inner edge to the planetesimal disk at $\sim0.5-0.7$~AU in order to reproduce Mercury's small mass \citep{Chambers2001,Hansen2009} and because the angular momentum of the terrestrial planets is inconsistent with accreting significant mass from interior to Venus' orbit \citep{Wetherill1978}.  Historically this was not viewed as troubling, because chemical condensation modeling \citep{Lewis1974} indicated that temperatures closer to the Sun would rise above the condensation temperature of any solids; these models' high temperatures at 0.5~AU also seemed to explain Mercury's metal-rich rich nature.  

However, 'dead zones' close to stars may inhibit MHD turbulence \citep{Lyra2009}, reducing energy dissipation and temperatures in these optically thick regions; planet formation may sequester dust rapidly \citep{Dzyurkevich2010} resulting in the STIPs regions being undetectable in the protostellar SED.   Other, ubiquitous disk processes also promote the rapid formation of planetary building blocks very close to the star \citep{Boley2014}.  Given a supply of solids interior to 1 AU, accretion simulations show planets forming easily in these regions \citep{Hansen2013}. Starting with the hypothesis that nearly all FGK stars form with a STIP, it is probable that such systems are dynamically metastable on a variety of timescales, allowing for planetary consolidation or destruction.  In this context, our explanation for the Solar System's current lack of large planets interior to $\sim~0.7$ AU is that our Solar System originally formed with a STIP at $<0.5$~AU composed of a few, now absent, Earth-scale planets.  Through multiple generations of catastrophic collisions and re-accumulations initiated by a dynamical instability between the original planets, we now have Mercury as the last remaining relic.

\section{Metastability in Planetary Systems}\label{s:stability}

Kepler host stars are $\sim$1--10~Gyr old \citep{Marcy2014}; obviously STIP formation must allow long-term dynamical stability, even if some of the systems' planets are nested at intervals barely beyond the stability requirements \citep{Lissauer2011}. Some well-studied systems are today on the edge of dynamical instability \citep{Lissauer2013,Deck2012,Lissauer2012}.  This seems completely reasonable: planet formation gradually combines dynamically unstable protoplanets, so evolving systems will rarely transition from being `highly coupled' to `extremely overstable'.  Thus planetary systems should naturally always be in a state of metastability \citep{Laskar1996}.  Our Solar System itself is only metastable; the terrestrial planets' orbits are chaotic \citep{Laskar1989}, and Mercury has a 1~\% chance of creating large-scale chaos on 5~Gyr timescales \citep{Laskar1996,Laskar2009}.  STIPs should exhibit similar metastability, with many systems metastable on the lifetime of their star, while others reached orbit crossing in the past. In this framework, the STIPs frequency found by Kepler represents a lower limit on their formation probability because we only see the Gyr stable systems.  The absence of STIPs around many stars could be due to the earlier collapse of a metastable planetary arrangement, leaving one or no detectable planets at short periods.

To explore metastable states in STIPs, we preformed a large suite of numerical integrations based on the observed, presumably Gyr stable, Kepler STIPs.  We generated systems with architectures {\it similar} to the known systems by using the observed semimajor axes and planetary radii, calculating planetary masses using the relationship $M_p~\simeq~M_E~(R_p/R_E)^{2.06}$ \citep{Lissauer2011}.  We randomized the initial orbital angles, assumed nearly coplanar orbits (mutual $i<1.5^\circ$), and assigned random initial eccentricities $e_0=$0--0.05; if the observed systems had measured maximal $e$, we assigned $e_0=$0--$e_{max}$.  The real Kepler systems' eccentricities are weakly observationally constrained (if at all), but the range we consider is consistent with the observations \citep{Fabrycky2014}.  We found that the range of $e_0$ for the architectures we explored was unimportant; instability probability is not correlated with $e_0$ (within our chosen range).

We integrated analogs of 13 observed Kepler systems with more than 4 planets: 
Kepler-102, Kepler-84, Kepler-90 \citep{Lissauer2014};  
Kepler-107, Kepler-169, Kepler-292, Kepler-223, Kepler-26 \citep{Rowe2014}; 
Kepler-11 \citep{Lissauer2013}; 
Kepler-62 \citep{Borucki2013}; 
Kepler-85 \citep{Ming2013}; 
Kepler-20 \citep{Gautier2012}; 
and Kepler-33 \citep{Lissauer2012}.  
These analogs were evolved using the Mercury orbital integrator \citep{Chambers1999} with an approximate inclusion of general relativity \citep{Lissauer2011}.  We integrated 20 analogs of each Kepler STIP for 10 Myr or until a physical collision occurred between two planets. For Kepler-169, 292, and 84, no analogs had a collision within 10 Myr from our initial conditions; additional 10-Myr simulations with $e_0=$0.05--0.1 also produced no collisions, indicating that these system architectures are likely stable on long timescales. In our framework these may represent systems which have had a planetary consolidation in the past, or which are simply unstable on longer, 1--10~Gyr timescales.  This does not imply that the other ten observed systems are currently unstable on 10 Myr timescales. We are {\it not} investigating the stability of the exact observed systems, which have been shown to be stable for $\sim$10$^8$~yr assuming initially circular orbits~\citep{Lissauer2011,Fabrycky2014}. We are instead interested in the range of possible behavior for STIPs with planet masses and spacings similar to the observed STIPs.

\begin{figure}[htbp]
\centering
\includegraphics[width=5.5in]{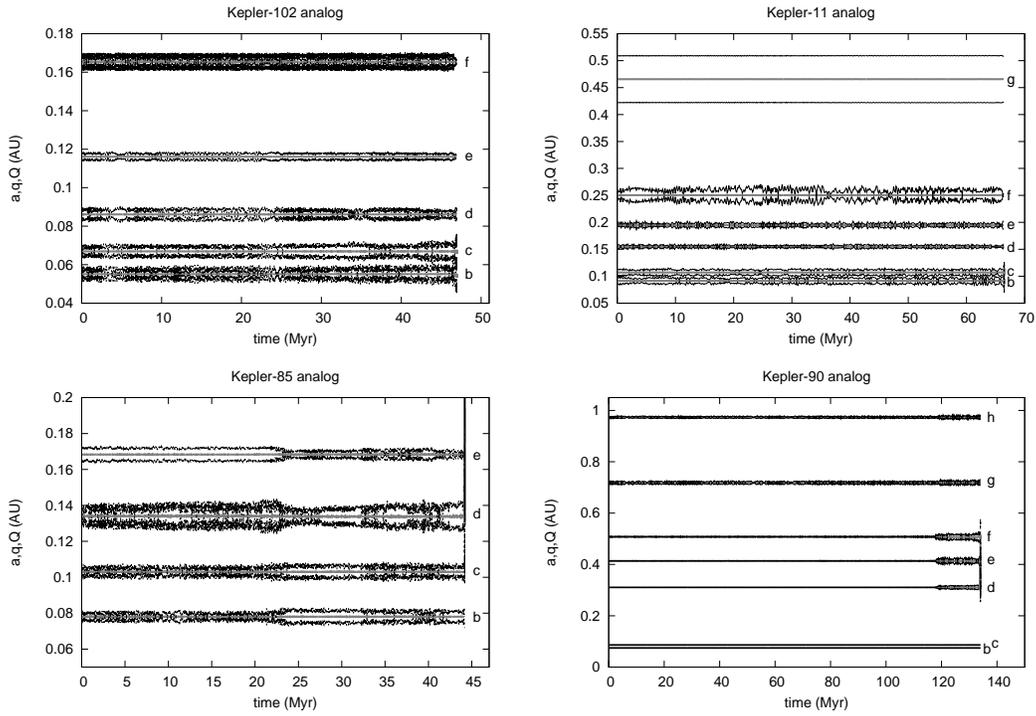}
\caption{Evolution of semimajor axis, periapse distance, and apoapse distance for four STIPs analogs.  Sometimes the innermost two planets collide leaving the outer planets relatively undisturbed (top panels) while other systems show close encounters between three planets (bottom panels).}
\label{f:ex-sys}
\end{figure}

For the ten systems that had collisions, we ran large suites of similar integrations for 100--150 Myr to explore the questions: (1) how do systems evolve as instability is approached? and (2) how are the instability timescales distributed? Fig.~\ref{f:ex-sys} shows typical evolutions for analogs which exhibited a first planetary collision after 45--140~Myr of metastability. Out of the $\sim$600 system analogs we integrated, half resulted in a collision between planets. What may be surprising is that there is no obvious sign of coming instability; as Fig.~\ref{f:ex-sys} shows, the systems maintain $e\approx~e_0$ for nearly the entire duration before suddenly transitioning to orbit crossing. Initial planetary eccentricities have no systematic effect on {\it when} the instability occurs, but even tiny differences in initial conditions can produce vastly different instability timescales.  Potential sources of chaos in exoplanet systems have recently been discussed by~\citet{Quillen2011} and~\citet{Deck2013}; how systems can evolve to a suddenly more chaotic state at an unpredictable time are discussed by \citet{Laskar1996}, \citet{Lithwick2014}, and~\citet{Batygin2015}.

\begin{figure}[htbp]
\centering
\includegraphics[width=5in]{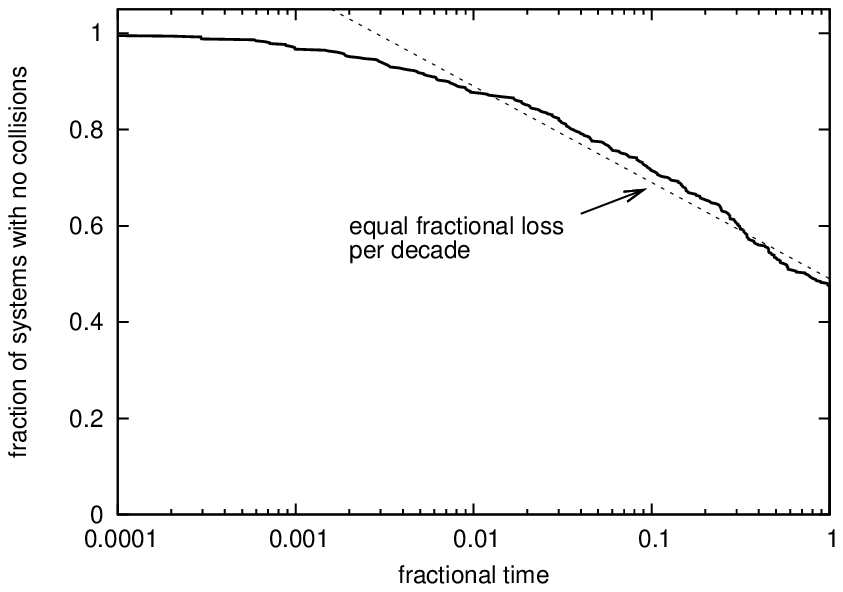}
\caption{Fraction of STIP analogs that have not experienced a collision as a function of integration time.  The dashed line shows a slope of 20\% loss per decade in time.}
\label{f:decay}
\end{figure}

Our experiments show that instability timescales in these systems are distributed such that equal fractions of the systems go unstable (reach a first planetary collision) in each decade in time (Fig.~\ref{f:decay}). This logarithmic decay is not unknown in dynamical systems ({\it eg.,}~\citet{Holman1993}) and is presumably related to chaotic diffusion.  After a brief, relatively stable initial period, the systems hit instability at a rate of $\sim$20\% per time decade, with half of the systems still intact at $\sim$100~Myr. The exact decay rate may be influenced by our usage of the current Kepler STIPs (perhaps the most stable); however if this decay rate held, at $\sim$5~Gyr 5--10\% of STIPs would not yet have reached an instability, in rough agreement with the observed STIPs frequency.

Mercury being a remnant of a previously unstable system fits nicely with its current instability timescale \citep{Lithwick2014}; it would also not be the first suggestion of a metastable Solar System configuration \citep{Gomes2005}. If our Solar System once contained a metastable set of planets interior to Venus, one can eliminate the artificial inner disk edge used in terrestrial planet formation models; the difficulty in making Mercury analogs \citep{Chambers2001,Hansen2009} would then be explained by the fact that Mercury is a collisional remnant.  Test integrations determined that the orbital evolution of the three outer terrestrial planets (Venus, Earth and Mars) are unaffected on 500 Myr timescales by the presence of four additional planets totaling 4$M_\earth$ in mass with $a\leq0.5$~AU. We also observe in our STIP simulations that when inner planets experience instability, the outermost planet (Venus analogs at $\gtrsim$~0.5~AU) is often unperturbed. Thus it is not unreasonable that Venus, Earth, and Mars could escape large-scale orbital effects as the Solar System's STIP disintegrates.

\section{Consolidation and/or Destruction}\label{s:destruction}

Once instability is initiated, the possible end-states of STIPs will fall along a continuum with two extremes: 1) {\it consolidation}, where almost all of the initial mass ends up in a smaller number of planets,  or 2) {\it destruction}, when $<$10\% of the STIP's  mass survives. We propose that our Solar System reached the destructive end state, but other systems consolidated an initial many planet STIP into fewer, more massive short-period planets.  The destructive end state is made possible by the extreme collision speeds which can occur for such close-in orbits; further from the star, the ratio of typical impact speeds, $v_{imp}$,  to mutual escape speeds, $v_{esc}~=~\sqrt{2G(M_1+M_2)/(R_1+R_2)}$, are low enough that accretional/consolidational processes are more likely. We note that most of the literature on planet-scale collisions has understandably focused on the context of accretion near 1~AU; at $\sim$0.1~AU impact speeds rise by factors of $\sim$3, greatly increasing the likelihood of erosive collisions.

\begin{figure}[htbp]
\centering
\includegraphics[width=5.5in]{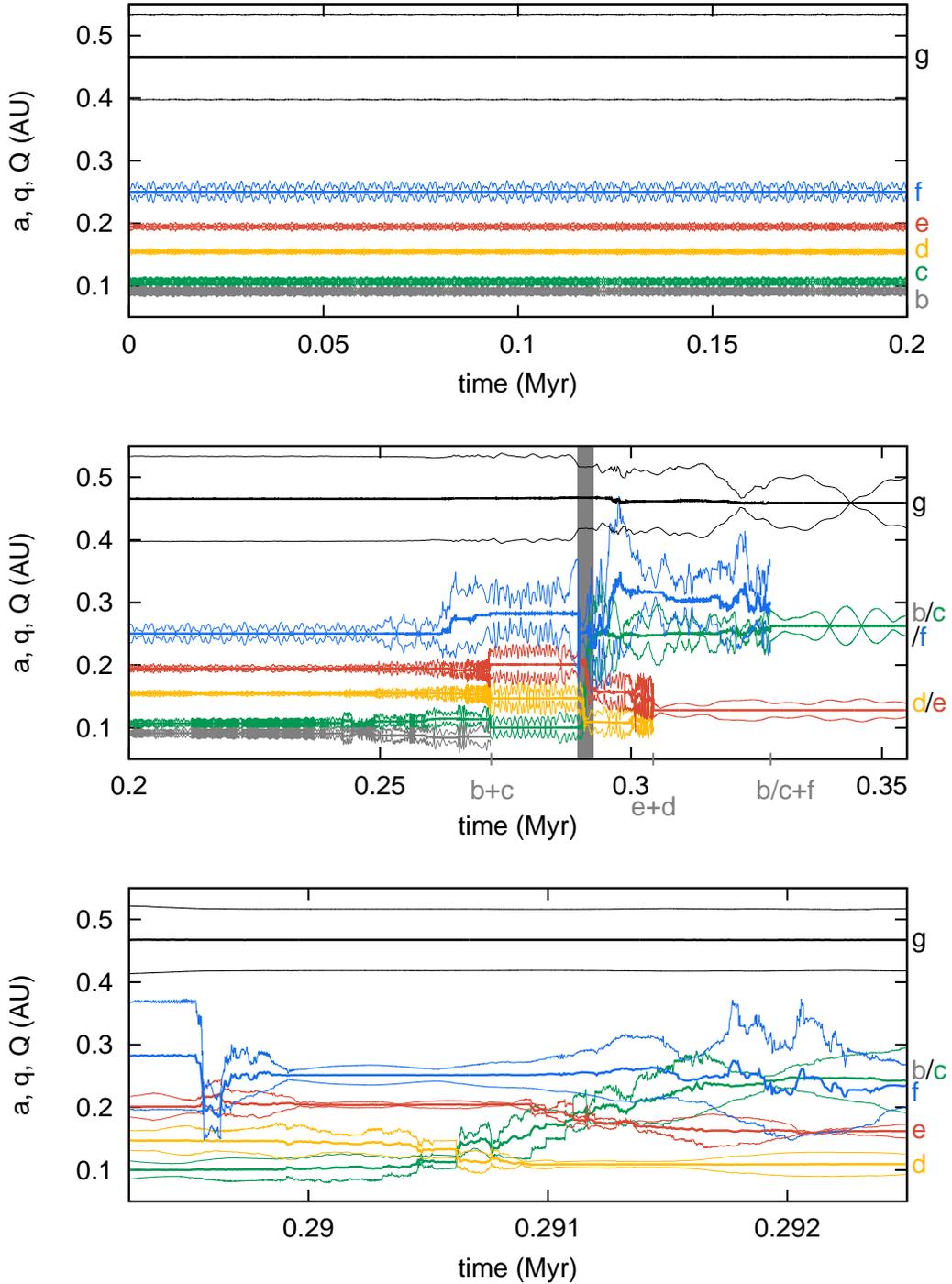}
\caption{Instability in a Kepler-11 analog.  An initial period of apparent stability (upper panel) transitions to a perturbed state with three collisional events (modeled as perfect consolidations) before reaching a final, 3-planet metastable state (middle panel); the shaded period between the first and second collisions involves four planets in dramatic `orbit swapping' (bottom panel).}
\label{f:k11}
\end{figure}

Fig.~\ref{f:k11} shows an example of how a system might evolve after an instability. Like in Fig.~\ref{f:ex-sys}, there is initially no macroscopic evidence of instability.  In the last 10\% of that phase, a transition occurs where several planets begin to interact, leading planets b and c to collide.  We had (incorrectly) postulated that essentially all analogs would reach their first collision as a result of an instability between only one pair of planets; we find instead that $\sim25$\% of the analogs show close encounters between 3 or more planets before the first pairwise collision occurs.

We continued some simulations beyond the first collision to study generic features of the subsequent evolution.  To do this, we assumed perfect, inelastic merging of the colliding planets (which is unlikely to be a good approximation for reasons discussed below).  The outcomes subsequent to the first collision are highly ergodic; in Fig~\ref{f:k11}, the 6 planet system consolidates to 3 planets which then remain stable for at least 100 Myr.  

\citet{Marcus2009} showed that the ratio $(v_{imp}/v_{esc}$) is important in determining the frequency of erosive collisions, which are more common when $(v_{imp}/v_{esc})>2$. Similarly, \cite{Stewart2012} showed collision outcomes as a function of  $(v_{imp}/v_{esc})$ and impact angle broken down into accretion, catastrophic collision, and `hit and run collisions' \citep{Asphaug2006} in which grazing collisions liberate some mass.  Even outside the catastrophic disruption regime, these authors point out that significant fractions ($\sim$10--15\%) of impacting mass in any collision is likely dispersed into small debris. 

\begin{figure}[htbp]
\centering
\includegraphics[width=5.5in]{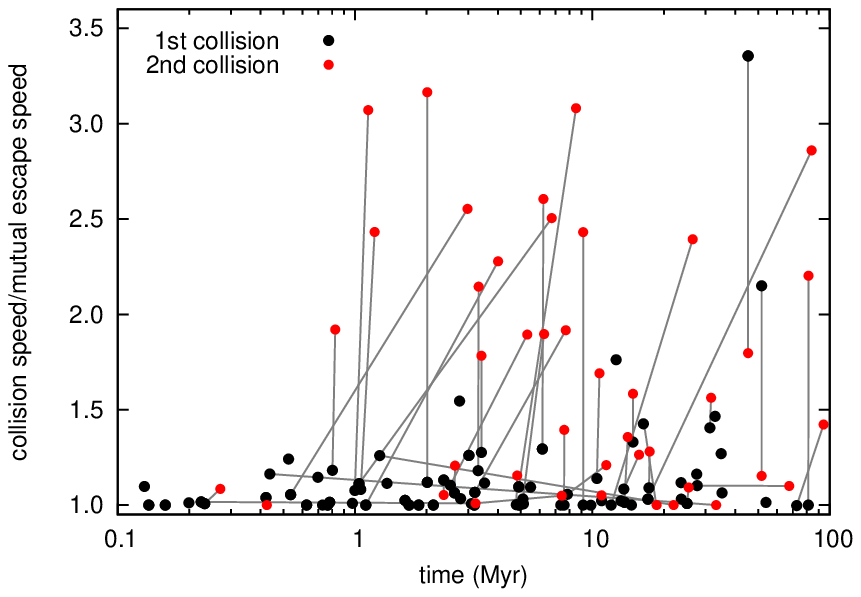}
\caption{Ratio of collision speeds to mutual escape speed vs time for STIP analogs. First collisions (black dots) are dominantly accretional ($v_{imp}/v_{esc}<2$). The second collision in a given system (red dots connected to the first event by a line) clearly shows that sequences of collisions allow more erosive conditions. The highest collision speed was 75~km/s}
\label{f:speed}
\end{figure}

We compiled $(v_{imp}/v_{esc}$) values from our simulations (Fig.~\ref{f:speed}); for systems initially spaced by 10--30 mutual Hill radii (like the Kepler analogs), if two comparable-mass neighboring planets at 0.1 AU are excited to mutual crossing, then $e\sim0.2$ at the time of first crossing, yielding encounter speeds $v_{enc}\sim~e~v_{kepler}\sim20$~km/s.  Super-Earths with $\sim1.6~R_\oplus$ have escape speeds near this value, so when the first pair of planets collide $v_{imp}=\sqrt{v_{enc}^2+v_{esc}^2}\sim~1$, as Fig.~\ref{f:speed} confirms.  These first collisions are likely  consolidational. If this was the whole story, one might think that STIPs gradually pairwise combine in a dominantly accretional environment.  However, our simulations show that {\it~subsequent} collisions often occur at much higher impact-speed ratios (Fig.~\ref{f:speed}).  The previous excitation in the system results in higher speed second collision, either nearly immediately (vertical lines in Fig.~\ref{f:speed}) or after a metastable delay (diagonal lines). Many of these second collisions rise into the more erosive regime; we propose multi-collision excitation is how some systems enter the destructive regime.

\section{Mass loss mechanisms}\label{s:massloss}

After instability, many STIPs may experience only a sequence of low-speed, accretional impacts between nearest-neighbor planets that produce small amounts of rapidly-eliminated debris. Such systems consolidate down to fewer planets, concentrating the mass spectrum toward larger planets.  Indeed the distribution of planet sizes in the Kepler systems with 3 or more planets show a trend toward larger planets at lower multiplicity (not formally significant).  Even in consolidational systems, instabilities can produce moderate inclinations, causing some surviving planets to become undetectable in transit surveys \citep{Johansen2012}. 

Our conclusion that some fraction of STIPs reach a first instability and then consolidate or degrade is independent of the hypotheses that follow regarding possible significant mass loss.  However, some of our STIPs analogs experience heavy perturbation; such systems are candidates for substantial mass loss either through continuous bursts of debris dispersal or even large-scale planetary elimination. We divide this into four size scales:
\begin{enumerate}
\item Dust below the blow-out limit ($\sim$1--10 microns) hyperbolically leaves the STIP region in just months.  This mass-loss mechanism is very efficient if planetary-scale events directly generate large amounts of dust or if dust is produced by a cascade of smaller collisions in the aftermath of each major event.

\item PR drag can cause cm-scale particles to spiral from 0.1~AU down to the star on timescales of 100 kyr.  However, if a collisional cascade produces a considerable amount of small debris, then the timescale for debris self-collision is shorter than the PR drag timescale \citep{Melis2012} and particles cannot inspiral before being reduced to dust and blown out.  
\end{enumerate}

\noindent The above two processes might be inefficient if $\sim$0.1~$M_\oplus$ of cm or smaller debris is produced in any single event because the optical depth to the star exceeds unity and the disk could self-shield \citep{Gladman2009}, shutting down radiation effects.  A competition can occur between the timescale for the largest remnant to sweep up debris and the timescale for debris to self-collide and grind down to the PR and dust blow-out scales. Moderate events ($\sim100$~km scale) in that collisional cascade produce sudden spikes in dust that quickly decay, perhaps like those observed by \citet{Meng2014,Song2005}.  Large-scale planetary violence finishes within $\sim5\times10^4$~yr (see Fig.~\ref{f:k11}), so only $~10^{-5}$ of mature field stars would show these sudden bursts of hot-dust excess. Because of the `equal fraction per decade' instability behavior, samples of younger stars would have hot-dust probabilities inversely proportional to their age.

\begin{enumerate}
\setcounter{enumi}{2}
\item For meter to 100 km objects, evolution is largely driven by repeated gravitational scatterings by the largest remnant(s).  With high relative orbital speeds, gravitational focussing (which enhances re-accretion) is minimized.  High $v_{imp}$ results in erosive impacts for most impact angles \citep{Marcus2009}, potentially removing, rather than adding, mass from the remnants. The 50--100 km/s impact speeds occurring this close to the star may enhance vapor production, hindering ejecta retention relative to slower impacts out near 1 AU.  In the Solar System, debris interior to $\sim0.5$~AU has very short collisional lifetimes and is subject to removal via Yarkovsky drift, consistent with the current lack of km-scale and larger debris in this region \citep{Vokrouhlicky2000}.

\item Close to a star, pure ejection is unlikely because $v_{esc}/v_{kepler}$ is so small.  However, secular interactions in a post-instability system could push planets or debris to star-grazing ($e\sim1$) on Myr timescales.  Although stellar impacts are unobserved in our simulations, $e$ pumping via secular resonances is a known phenomenon in our Solar System \citep{Namouni1999,Laskar2009} and often relies on the presence of exterior giant planets which are unseen (and thus unmodeled) in the Kepler systems. While the Kepler planets are obviously not {\it today} near secular resonances, post-instability evolution could change this; Fig.~\ref{f:k11}c illustrates scattering planets exploring a large range of $a,e,i$ space, a near-perfect algorithm to find secular resonances.  It is difficult to assess the probability of significant secular evolution, but we ran simulations of pairs of $\sim$Earth-mass planets evolving in our current Solar System (minus Mercury) on low-$e$ orbits form $0.1-0.4$ AU; this configuration is a plausible outcome of a recently consolidated 3-planet STIP. We find many configurations where these planets' eccentricities grow to $e=0.4-0.9$ on $\sim1-10$~Myr timescales, with the inner planet's perihelion distance sometimes dropping to just a few Solar radii.  Although achieving $e\sim1$ is rare, moderate secular eccentricity growth can promote subsequent high-speed collisions between the remaining planets.

\end{enumerate}

We thus postulate that our Solar System originally had a STIP of 3 or more now-absent planets totaling a few Earth masses.  An instability initiated a sequence of collisions (as opposed to a single collision of a $\sim0.2M_{\oplus}$ body, \citet{Benz1988}), which allowed the system's excitation to the destructive regime; such a process concentrates iron into the surviving remnants, explaining Mercury's high density \citep{Stewart2012,Asphaug2014}.   Mercury's current $e\sim~i$~(radians)~$\sim~0.2$ \footnote{That is, $e\sim~v_{esc}/v_{kep}\sim$~(20~km/s)/(100~km/s)} and Mercury's current `survivor' metastability timescale of $\sim$5~Gyr would naturally result from this scenario.  Additionally, the transplant of iron-meteorite parent bodies to the asteroid belt \citep{Bottke2006} from the $<~0.5$~AU region in this scenario can easily accommodate their rapid initial accretion and evidence for grazing protoplanetary impacts \citep{Goldstein2009}.

Much work beyond this Letter is required to explore the details of this general framework.  For example, the probability of destructive end states should be consistent with the de-biased estimate that half of mature FGK stars have no visible planet with $R>1~R_{\oplus}$ and $P<88$~days \citep{Fressin2013}.

\section{Mercury and the Lunar Cataclysm}\label{s:lc}

The decadal nature of the instability makes it plausible that the penultimate metastable state lasted $\approx0.5$~Gyr (one tenth Mercury's current metastablility timescale), allowing the additional speculation that this last instability, 4 Gyr ago, was responsible for the ``Lunar Cataclysm'' (reviewed by \citet{Hartmann2000}).  An instability transitioning into rapid planetary destruction (in $\sim$1~Myr) would spread debris throughout the inner Solar System. \citet{Gladman2009} estimated that 10--20\% of m to 100~km debris originating near current Mercury would strike Venus, with 1--4\% impacting Earth ($\sim$0.1\% strikes the Moon).  The Earth's impact rate would peak $\sim$1--10~Myr after the event and decay on $\sim$30~Myr timescales as Mercury and Venus absorb most of the debris; this is a plausible match for the cataclysm's final stages \citep{Cuk2010}.  A bottom-heavy size distribution for the 1--100 km debris could explain the recent finding \citep{Minton2015} that a main-belt asteroid source would produce too many impact basins during the cataclysm.  STIP debris would likely be mostly silicate-rich mantle material similar but not identical to main-belt asteroid compositions, consistent with cataclysm impactor compositions inferred via cosmochemical means \citep{Joy2012}.  The smallest dust (being blown out hyperbolically) could impact the Earth-Moon system. We estimate that $10^{-11}$ of the departing dust would strike the Moon, at $v_{imp}\sim$30~km/s.  If any dust or meteoroid projectiles were retained, fragments might be found in regolith breccias compacted during the cataclysm epoch. Compared to traditional cataclysm hypotheses, this scenario yields significantly higher impact rates onto Venus, with potentially significant implications for its evolution.

\noindent{\bf Acknowledgements}: We would like to acknowledge helpful discussions with Aaron Boley, Norm Murray, Scott Tremaine, Brenda Mathews, Doug Hamilton, and Sarah Stewart.  We thank the Natural Sciences and Engineering Research Council of Canada and the Canadian Institute for Theoretical Astrophysics for financial support.

\clearpage

\end{document}